\documentstyle[12pt,fps,cite,oldlfont]{article}

\makeatletter
\@addtoreset{equation}{section}
\makeatother

\newcommand{\eq}{\ref}
\newcommand{\beq}{\begin{equation}}
\newcommand{\eeq}{\end{equation}}
\newcommand{\bea}{\begin{eqnarray}}
\newcommand{\eea}{\end{eqnarray}}

\newcommand{\cc}{\cite}
\newcommand{\lb}{\label}

\newcommand{\ul}{\underline}

\newcommand{\chup}{$\ch U\ph$}
\newcommand{\scsb}{S$\chi$SB}

\newcommand{\pdc}{\ph^{\dagger}\ch}

\newcommand{\cbcex}{\langle\chb \chi \rangle}
\newcommand{\cbc}{\chb \chi}
\newcommand{\phex}{\langle\ph\rangle}

\newcommand{\chb}{\overline{\chi}}

\newcommand{\SG}{SGM}

\newcommand{\ds}{\displaystyle}

\newcommand{\half}{\frac{1}{2}}

\newcommand{\bt}{\beta}
\newcommand{\lag}{\langle}
\newcommand{\rag}{\rangle}
\newcommand{\gm}{\gamma}

\newcommand{\vep}{\varepsilon}

\newcommand{\kp}{\kappa}

\newcommand{\sg}{\sigma}

\newcommand{\ph}{\phi}

\newcommand{\ch}{\chi}
\newcommand{\ps}{\psi}

\newcommand{\Om}{\Omega}

\newcommand{\psb}{\overline{\ps}}

\newcommand{\ra}{\rightarrow}

\newcommand{\be}{\begin{equation}}
\newcommand{\ee}{\end{equation}}

\def \3{\ss}

\def\dateandnumber(#1)#2#3#4{
\vbox to 18mm{%
     \hbox to \textwidth{ \hspace*{14mm} \hsize=40mm%
            \vbox{%
                 \hbox to 40mm{\large #1 \hss}%
                 \hbox to 40mm{    \hss}%
                 \hbox to 40mm{    \hss}%
                 }%
                 \hss \hsize=80mm%
            \vbox{%
                 \hbox to 80mm{\hss \large #2}
                 \hbox to 80mm{\hss \large #3}
                 \hbox to 80mm{\hss \large #4}
                 }%
            \hspace*{14mm} }%
      \vss
    }
}
\def\titleofpreprint#1#2#3#4{{\LARGE \bf
\vbox to 43mm{%
     \vss
     \hbox to \textwidth{ \hspace*{14mm} \hsize=130mm%
            \hss \vbox{
                      \hbox to 130mm{\hss \LARGE \bf #1\hss}%
                      \hbox to 130mm{\hss \LARGE \bf #2\hss}%
                      \hbox to 130mm{\hss \LARGE \bf #3\hss}%
                      \hbox to 130mm{\hss \LARGE \bf #4\hss}%
                 }%
            \hss \hspace*{14mm} }%
      \vss
    }
}}

\def\listofauthors#1#2#3{{\large
\vbox to 22mm{%
     \vss
     \hbox to \textwidth{ \hspace*{14mm} \hsize=130mm%
            \hss \vbox{
                      \hbox to 130mm{\hss \large #1\hss}%
                      \hbox to 130mm{\hss \large #2\hss}%
                      \hbox to 130mm{\hss \large #3\hss}%
                 }%
            \hss \hspace*{14mm} }%
      \vss
    }}
}
\def\listofaddresses#1#2#3#4{{\small
   \vbox to 18mm{%
        \vss
        \hbox to \textwidth{ \hspace*{14mm} \hsize=130mm%
               \hss \vbox{
                         \hbox to 130mm{\hss \small #1\hss}%
                         \hbox to 130mm{\hss \small #2\hss}%
                         \hbox to 130mm{\hss \small #3\hss}%
                         \hbox to 130mm{\hss \small #4\hss}%
                         }%
               \hss \hspace*{14mm}
        }%
        \vss
   }}
}
\def\abstractofpreprint#1{{\normalsize
\vbox to 110mm{%
     \vss
     \hbox to \textwidth{\hss \normalsize \bf Abstract \hss}%
     \normalsize
     #1
     \vss
     }}
}
\def\footnoteoftitle#1{{\small
\vbox to 30mm{\parindent0pt
     \vss\small #1 \vss
    }}
}
\def\footnoteitem(#1)#2{
\begin{list}{#1}{\labelwidth4.0mm \leftmargin7.0mm
\labelsep2.5mm \rightmargin7.0mm \parsep0.5ex plus0.2ex minus0.1ex
\itemsep0ex plus0.2ex }
\item #2
\end{list}
}
\begin{document}
\dateandnumber(November 1994)%
{J\"ulich, HLRZ 52/94}%
{                    }%
{                    }%
\titleofpreprint%
{            Dynamical fermion mass generation                 }%
{            by strong gauge interaction                       }%
{              shielded by a scalar field$^*$                     }%
{                                                              }%
{                                                              }%
\listofauthors%
{       Christoph~Frick$^1$ and Ji\v{r}\'{\i}~Jers\'ak$^2$     }%
{                                                              }%
{                                                              }%
\listofaddresses%
{\em Institute of Theoretical Physics E,                        %
     RWTH Aachen,         D-52056 Aachen, Germany              }%
{     and                                                       %
                                                               }%
{\em HLRZ c/o KFA J\"ulich,                                     %
      D-52425 J\"ulich, Germany                                }%
{                                                               %
                                                               }%

\abstractofpreprint{
The strongly coupled lattice gauge models with confined fermion and scalar matter fields, which in a certain phase break dynamically a global chiral symmetry, are reconsidered from the point of view of the existence of heavy fermions.
If these models are interpreted as describing a new strong force beyond the standard model, such heavy fermions can arise as neutral bound states of the fundamental fermion and scalar.
We call this mechanism of dynamical fermion mass generation {\em shielded gauge mechanism}. 
The scalar field induces at strong gauge coupling a second order phase transition which is necessary for a continuum limit.
Therefore the mechanism might well exist also in continuum.
In this case, assuming that strongly coupled chiral gauge theories with scalars have similar dynamical properties at strong coupling as the vectorlike models investigated on the lattice, the discussed  mechanism could be considered as an alternative to the Higgs mechanism.
In particular, if the broken global chiral symmetry is SU(2) and the heavy fermion interpreted as top quark, the mechanism is analogous to some gauge models for top condensate.
We present some numerical data obtained in the quenched approximation of a model with vectorlike U(1) gauge symmetry.
The observed scaling behaviour of the chiral condensate and of the fermion mass and also the properties of Goldstone bosons are the first encouraging steps in a study of the continuum limit of the mechanism. 
}
\footnoteoftitle{
\footnoteitem($^*$){ \sloppy
Supported
by Deutsches Bundesministerium f\"ur Forschung und Technologie and
by Deu\-tsche Forschungsgemeinschaft.
}
\footnoteitem($^1$){ \sloppy
Present address: AMS, Querstr. 8-10, D-60322 Frankfurt a.M., Germany
}
\footnoteitem($^2$){ \sloppy
E-mail address: HKF204@zam001.zam.kfa-juelich.de
}
}
\pagebreak

%

\section{Introduction}
The search for alternatives to the Higgs mechanism generating the masses of gauge bosons and fermions in the electroweak theory is motivated by a feeling that this mechanism is hand-made, and some dynamical explanation might be actually possible (for a recent review see ref.~\cc{Ba93}). 
A fundamental Higgs field is considered to provide a comfortable description of the spontaneous symmetry breaking, allowing a perturbative approach, but its use might not be really necessary. 
Therefore most alternative attempts try to avoid the introduction of a fundamental scalar field. 
The obligatory Goldstone bosons, as well as the Higgs boson, should it exist, are interpreted as composite states, formed by some strong dynamics beyond the standard model. 

However, one should distinguish between a fundamental scalar field whose selfinteraction triggers the symmetry breaking already on the tree level, as in the conventional Higgs mechanism, and such a field playing some different important role in the dynamical symmetry breaking occuring only beyond the perturbative expansion.
In this second role a fundamental scalar field is acceptable provided its particle excitations are massive or confined. 

Already known examples are strongly coupled Yukawa models, which can exhibit spontaneous chiral symmetry breaking (\scsb) even if the bare scalar potential in the action  does not have the form of the classical Mexican hat~\cc{KoTa90,Ca91,ClRo91}. 
On the lattice the \scsb\ can occur in such models even if the nearest neighbour coupling of the scalar field is antiferromagnetic, i.e.~competing against the \scsb\ generated dynamically by the Yukawa interaction term (see e.g.~\cc{LeShi90a,BoDe92b}, for reviews see refs.~\cc{Shr92,DeJe92}).
The scalar boson can be made heavy within the joint upper bounds on the Higgs
and fermion masses~\cc{BoFr93a,BoFr93b,FrLi93,LiMu93}. 

In this paper we would like to point out that the experience accumulated on the lattice with strongly coupled vectorlike gauge theories with matter fields suggests the existence of still another alternative of dynamically generated \scsb. 
The mechanism we want to describe assumes some new confining gauge field~$A$ of a compact gauge group~$G$, and makes use of a fundamental scalar field~$\ph$ which is coupled to this gauge field and, consequentely, confined. 
The scalar field, however, does not generate the \scsb, in fact it is crucial that~$\phi$ acts against it. 
The \scsb\ is generated dynamically by the interaction between the gauge field~$A$ and some fermion fields~$\ch$, making the gauge invariant condensate~$\cbcex$ nonzero, in an analogy to the \scsb\ in QCD.

The role of the scalar field~$\ph$ is twofold. 
First, it shields the $G$-charge of the fermion $\ch$, so that composite $G$-neutral physical fermion states of the form~$F = \pdc$ can exist asymptotically in spite of the confinement of the $G$-charge. 
This is why we call our proposal the {\em shielded gauge mechanism} (SGM).
In the phase with \scsb\ ({\em Nambu phase}) the fermion mass~$m_{\rm F}$ is nonzero,
so we have the case of dynamical mass generation (DMG).
The Goldstone bosons are composed of $\chb$ and $\ch$.
We use the terms ``composite'' and ``dynamical'', whose meaning is somewhat obscure in strongly coupled field theories \cc{HaHa91}, essentially in the same sense as they are used in QCD. 

Second, to make the model applicable in the continuum physics, we have to make the lattice constant small and thus approach a phase transition of 2$^{\rm nd}$ order.
Usually phase transitions in the lattice gauge theories without scalars occur at  couplings of order one (lattice QED) or even at zero coupling (lattice QCD).
But the scalar field has the tendency to suppress the \scsb\ and induces for very strong gauge coupling a new second order phase transition, at which the chiral symmetry is smoothly restored.
In the scaling region of this phase transition in the Nambu phase, the mechanism  might be applicable to the continuum physics if the corresponding field theory is nonperturbatively renormalizable. 

What kind of continuum physics?
If e.g. the broken global chiral symmetry is SU(2) with one fermion field doublet $\ch$ coupled to~$A$, we find a massive fermion doublet $F= \pdc$ and three massless Goldstone bosons $\pi$, all being gauge invariant with respect to~$G$.
Because of confinement due to the new gauge interaction there are no physical states  corresponding to the field~$\ph$, but bosonic gauge invariant states consisting of~$\ch$,~$\ph$ and~$A$, e.g. of the type~$\ph^{\dagger}\ph$ and~$\cbc$, should be expected, some of them possibly looking like the Higgs boson.
The situation is then quite similar to the standard model with the standard gauge fields switched off, symmetry breaking present, and one degenerated weak isospin fermion doublet heavy. 
The other fermions, not coupling to $A$, are massless. 
If the standard SU(2)$\otimes$U(1) gauge fields are then switched on, the broken global SU(2) changes into the local one, and the $\pi$'s lead to the massive vector bosons (see e.g. \cc{LiRo92,Li93}).
The Higgs mechanism has been replaced by the SGM but, of course, the fermion spectrum is quite nonrealistic, except possible speculations about a very heavy fourth family with small mass differences in weak isospin doublets.

A more realistic application is suggested by ideas ascribing the top quark a special role in the symmetry breaking.
In fact, the mechanism we are describing might be related to the top quark condensate models based on new strong gauge interactions at some energies beyond the electroweak scale, as suggested by various authors~\cc{LiRo92,Li93,Ho87,Hi91,Boe91,Li92,Ma92}%
\footnote{We thank M. Lindner for many elucidating discussions on this idea},
and our suggestion is inspired by their work. 
Here it is e.g. assumed that the top-bottom doublet couples strongly to~$A$
whereas the other fermions couple only weakly or not at all to it, the interaction being flavour diagonal. 
The problem is that these models are chiral gauge theories which we still cannot simulate on the lattice.
Thus for this kind of physical application of the \SG\ we have to assume that the dynamics of strongly coupled chiral gauge theories with scalars is similar to the vectorlike lattice models.

We point out that the dynamical scenario we consider is different from the hypothetical strongly coupled standard model~\cc{AbFa81a,AbFa81b,ClFa86} searched for in vain some years ago within the same class of lattice models. 
There a confining phase with shielded fermions but without the chiral condensate has been assumed, whereas we are looking for a confining phase with both the shielded fermions and the nonvanishing chiral condensate.

On the lattice, such a phase is known since long, and the question is what are the properties of the continuum limit in this phase.
This is a fundamental question concerning the strongly coupled, not asymptotic free gauge theories.
The long term aim is to find out whether and how far the physical content of the SGM is different from the usual Higgs mechanism.
We do not insist on the existence of some new nontrivial fixed point, accepting the possibility that this mechanism could operate only in some energy range in analogy to the ``trivial'' Higgs-Yukawa sector of the standard model.
Even in such a case significant differences from the usual Higgs mechanism might be possible, as for example higher upper bounds for its validity.

To discuss the \SG\ we consider a specific chirally symmetric lattice  model with a staggered fermion field~$\ch$, the gauge field described by the group elements~$U$, and the complex scalar field~$\ph$ of fixed absolute value (\chup\ {\em model}).
As the gauge group we choose the compact~$G$ = U(1) group with both scalar and fermion fields having ``charge'' one with respect to it. 
This lattice model can be considered as a generic case for the models we have in mind, since it is a confining theory at strong coupling and thus similar also to nonabelian theories.
We note, however, that some similar extensions of the standard model based on the U(1) gauge group with strong coupling exist also in continuum \cc{LiRo92,Ho87,Boe91}.

In spite of the existing experience with this and similar lattice models, the proposed \SG\ is yet quite speculative because the critical line, which should be used for an approach to the continuum limit, is at present only poorly understood.
The scaling behaviour and the properties of the continuum limit at this transition are not yet known.
However, we argue that at least in the limit of infinite gauge coupling the SGM makes sense. 
Here we recover a lattice transcription of the
Nambu--Jona-Lasinio~(NJL)~model which has been used by various authors
as an effective theory for some strong dynamics with \scsb\ beyond the standard~model~\cc{Na89,MiTa89,BaHi90}.
This observation is based on the Lee-Shrock~transformation~\cc{LeShr87a}
relating the lattice theories with fermions, gauge and scalar fields
at infinite gauge coupling to a pure four fermion theory on the lattice (see sec.~2.3).
Thus it is possible that the SGM can be described by the NJL~model in the low~energy limit and that the \chup\ model is a generalization of the NJL model.

In this paper we report results of a nonperturbative investigation of the SGM by means of numerical simulations of the 4D \chup~models on the lattice in the quenched approximation.
Several \chup~models in~4D have already been studied numerically some time ago, both in the quenched approximation (e.g.~\cc{LeShi86d,LeShr87b,DeShi88,LeShr88b,Shr89}), and with dynamical fermions~\cc{DaKo88,AoLe88,Ku89a,DaMe89,MePe91}.
The aim was either to clarify the general features of the models or to search for the strongly coupled standard model.
But the questions we are asking now require a more systematic investigation of several relevant physical quantities in the Nambu phase,
in particular the chiral condensate, the fermion mass and the bosonic spectrum.
The reason is that we want to look for lines of constant mass ratios, which can best elucidate the physical content in the continuum limit.

Up to now we obtained quenched numerical results for the chiral condensate,
the fermion mass and the pseudoscalar Goldstone boson mass, and have got an insight about the behaviour of these quantities in the vicinity of the critical line, where the approach to the continuum limit should be performed. 
We also gained first experience with the determination of scalar and vector
bound states. 
These numerical results confirm the existence of the Nambu phase and the
expectations about the scaling of some fermionic observables.
They support the hope that a sensible continuum limit can be found.
For the long-term~goal to investigate the renormalizability properties
of the \chup~models in 4D further simulations with dynamical
fermions~\cc{FrFr95} are necessary.

We want to remark that the SGM can be considered also in lower dimensional models. 
As will be explained, the relationship between the \chup\ models and four fermion theories at strong gauge coupling \cc{LeShr87a} is independent of the dimension.
Therefore both in 3D and 2D the SGM exists as a renormalizable quantum field theory at least in the strong coupling limit of the \chup\ model. 
Namely, in 3D we  recover in this limit the nonperturbatively renormalizable~\cc{Gr75,RoWa89b,RoWa91} 3D~Gross-Neveu model. 
However, we do not yet know whether in 3D the SGM for a large, but finite, coupling is in the same universality class as this model.

In 2D, we find at infinite gauge coupling the cherished asymptotically free
chiral Gross-Neveu model \cc{GrNe74} which is thus related to the SGM in 2D at large coupling.
This 2D~model exhibits DMG without \scsb~\cc{Wi78b,DeFo93b}
due to the Mermin-Wagner-Coleman~theorem.
This shows that the DMG is a little bit more general phenomenon
than the \scsb\, and we therefore prefer to use the term DMG in this paper.

The outline of the paper is as follows: In the following section we review the already known relevant properties of the lattice \chup\ model we are investigating.
In section 3 we describe the properties of the Nambu phase and of its critical boundary.
Then we discuss the question of renormalizability. 
In section 4 the numerical data are presented.
In section 5 we give a summary and conclude with several speculative remarks.

\section{Lattice \chup~model at strong coupling}

Here we summarize the previous relevant knowledge about the lattice \chup\ models  with coupled fermion field~$\ch$, gauge field variable~$U$ and scalar field~$\ph$.

\subsection{The \chup~model with U(1) gauge symmetry}
%
For definiteness we make the following choices on the euclidean hypercubic lattice in~$d$ dimensions:
\begin{itemize}
\item~$\ch$ is one staggered fermion field. The model has the continuous global U(1) chiral symmetry with respect to the transformations
\be
   \chi_x \ra e^{i \alpha \vep_x} \chi_x \;, \hspace*{1.5cm}
   \chb_x \ra e^{i \alpha \vep_x} \chb_x \;, 
\lb{chtrafo}
\ee
with the standard~$\vep_x = (-1)^{x_1+ \cdots +x_d}$.
This is, however, only a residual symmetry, achieved on the lattice, whereas in the continuum limit the expected global chiral symmetry is U($N_f$)$\otimes$U($N_f$), $N_f$ being the number of fermion species found in this limit.
The charge of~$\ch$, determining its U(1) gauge transformation properties, is one.
\item $U$ is the gauge field defined on the lattice links.
The link variables are elements of the compact gauge group U(1).  
They can be imagined as
\be
   U_{x,\mu} = e^{i g a A_{\mu} (x)}  \; ,
\ee
with~$A_\mu(x)$ being the abelian gauge field,~$g$ the gauge coupling constant and~$a$ the lattice constant.
\item $\ph$ is a complex scalar field of charge one. 
It is convenient to impose the constraint~$|\ph |$=1, which corresponds to the choice of infinite quartic scalar selfcoupling.
This constraint is known not to restrict the physical content of the scalar sector.
For example block spin transformations set this constraint off.
\end{itemize}

The action is
\be
    S_{\ch U \ph} = S_\ch + S_U + S_\ph \; ,
\lb{action}
\ee
where
\bea
  S_\ch &=& {\textstyle \half} \sum_x \chb_x \sum_{\mu = 1}^d 
             \eta_{\mu x} \left[ U_{x,\mu} \chi_{x + \mu} -
                     U_{x-\mu,\mu}^\dagger  \chi_{x - \mu} \right] 
              \, + \, a m_0 \sum_x \chb_x \chi_x \; , 
\lb{SCH}                  \\
  S_U   &=& \bt \, \sum_{\rm P} \left[ 1 - \mbox{Re} \{ U_{\rm P} \} \right] \; , 
\lb{SU}  \\
  S_\ph &=& - \kp \, \sum_x \sum_{\mu=1}^d \left[ \phi_x^\dagger
              U_{x,\mu} \phi_{x + \mu} \,+\, {\rm h.c.} \right] \; . 
\lb{SPH}
\eea
Here~$\bt = a^{d-4}/g^2$, $U_{\rm P}$ is the plaquette product of link variables $U_{x,\mu}$ and $\eta_{\mu x} = (-1)^{x_1 + \cdots + x_{\mu - 1}}$.
All the fields and coupling parameters are dimensionless and therefore the calculated masses and expectation values are always in 
the lattice units, e.g.\ the masses are of the form~$am$,~$m$ being
the mass in physical units.
The hopping parameter~$\kp$ vanishes (is infinite) when the squared bare scalar mass is positive (negative) infinite.
The bare fermion~mass~$m_0$ is introduced for technical reasons, and the model is meant in the limit~$m_0 \!=\! 0$. 

\subsection{Properties of various subsystems in 4D} \label{subsystems}
\begin{figure}
\centerline{                                                                    
\fpsxsize=13.0cm                                                                 
\fpsbox{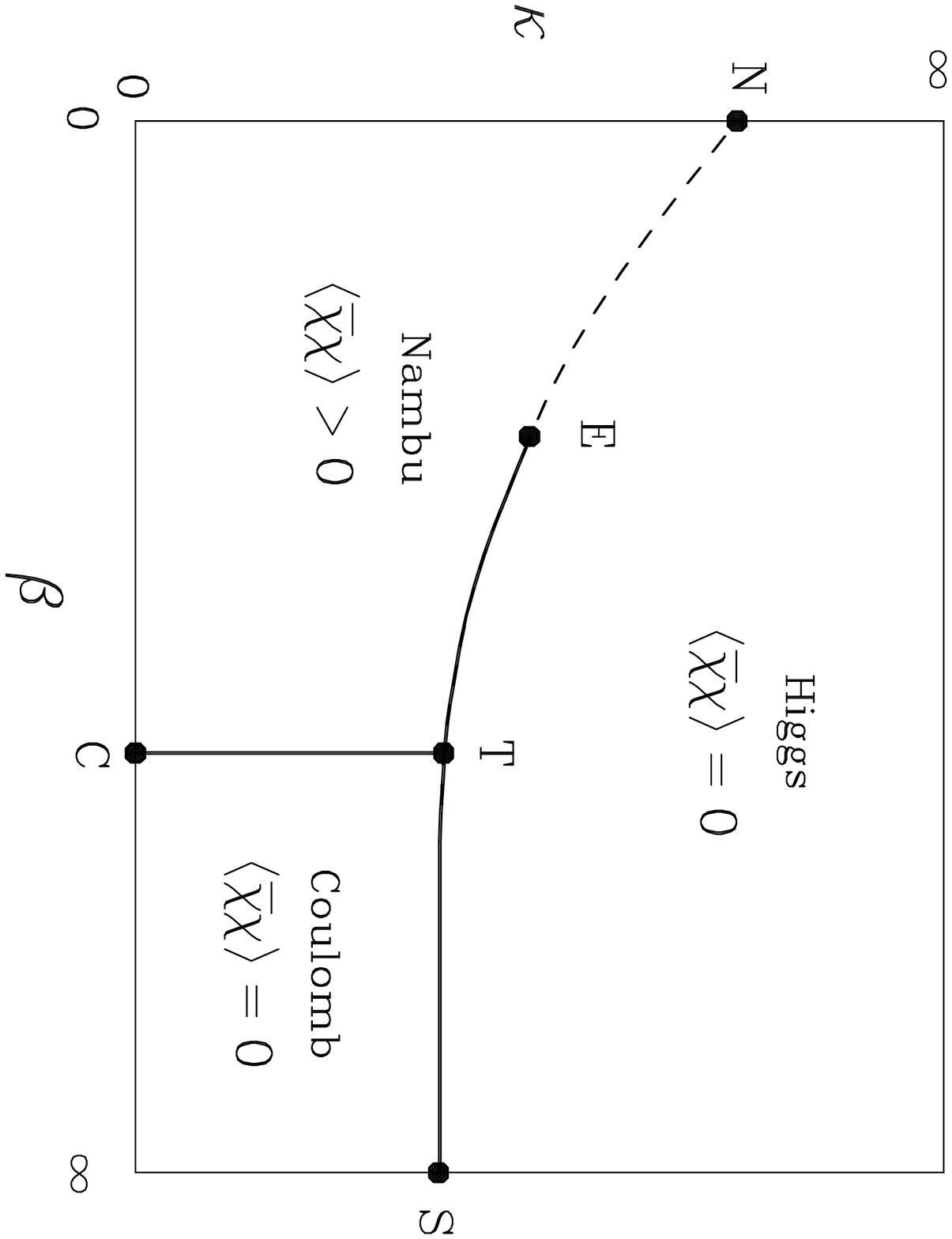}                                                               
}                                                                               
\vspace{-1.cm}                                                                 
\caption%
[...caption in list of figures...]%
{ 
The schematic phase diagram of the~\chup~model~({\protect\eq{action}}) in~4D at $m_0 = 0$.
The emphasized points are: \\*[3mm]
\phantom{~~~}{\bf N}:~ critical point of the {\bf N}JL~model, which is the limit 
        of the~\chup\ model at~$\bt \!=\! 0$, \\
\phantom{~~~}{\bf E}:~ critical {\bf E}ndpoint of the 
           Higgs phase transition line~ETS, \\
\phantom{~~~}{\bf T}:~ {\bf T}riple point, \\
\phantom{~~~}{\bf C}:~ phase transition from the {\bf C}onfinement (at strong gauge coupling)
                 to \\
\phantom{~~~{\bf C}:}~ the {\bf C}oulomb phase (at weak gauge coupling) 
                 in the model without the scalar field. \\
\phantom{~~~}{\bf S}:~ critical point of the {\bf S}pin model.
\\*[3mm]
The Nambu, Higgs and Coulomb phases are described in the text.
The dashed line corresponds to a 2$^{\rm nd}$ order phase~transition,
full lines to 1$^{\rm st}$ order transitions. 
}
\label{PD}                                                                   
\end{figure}

%
First we shall describe those properties of various subsystems which are relevant for the understanding of the whole \chup\ model in 4D at strong gauge coupling and of its phase diagram shown schematically in fig.~\ref{PD}.

%

\begin{itemize}
\item \ul{$S_\ch$ at~$g\!=\!0$:} Setting~$U_{x,\mu}\!=\!1$, this action describes a free staggered fermion field. 
Because of the lattice fermion doubling 
the model in 4D describes $N_f \!=\! 4$ fermions in the scaling region.
For $m_0 = 0$, the global chiral symmetry is then U(4)$\otimes$U(4). 
\item \ul{$S_U$:} This is the pure gauge theory on the lattice.
For a compact gauge group it is confining at small~$\bt$ (strong coupling) in the sense of the Wilson~loop criterion. 
The chosen U(1) gauge theory with action (\ref{SU}) is a generic example. 
It has a phase transition around~$\bt \!=\! \bt_c \simeq 1.0$, separating the confinement and Coulomb phases. 
The spectrum in the confinement phase consists of glueball-like states (``gaugeballs'').
\item \ul{$(S_\ch + S_U)$:} The coupled fermion-gauge system exhibits for small~$\bt$ the \scsb\ with a nonvanishing chiral condensate~$\cbcex$ and massless Goldstone~bosons. 
The system is also confining in the sense that only gauge invariant states like~$\cbc$ and gaugeballs are present in the spectrum. 
The situation is quite similar to QCD. 
The confinement-Coulomb phase transition is of first order~\cc{KoDa87} and occurs at~$\bt \!= \bt_c \simeq 0.89$ for~$N_f \!=\!4$ (point~C in fig.~\ref{PD}). 
\item \ul{$S_\ph$ at~$g\!=\!0$:} We restrict our description to~$\kp \ge 0$. 
The pure scalar field theory, selfinteracting in consequence of the constraint~$|\ph| = 1$, has the global U(1) symmetry, i.e.\ it is the XY~model in 4D.
This symmetry is gauged for~$g > 0$. 
At $g = 0$ it is spontaneously broken for~$\kp > \kp_c$ with~$\kp_c \simeq 0.15$.
Below~$\kp_c$ the symmetry is restored. 
The phase transition at~$\kp_c$ (point~S in fig.~\ref{PD}) is of magnetic type, the order parameter being~$\phex$ as in spin models of ferromagnetism.
The mass~$am_\ph$ of the lowest mass excitation of the~$\ph$ field below~$\kp_c$, the~$\ph$ boson doublet, diverges as~$\kp \ra 0$ and vanishes as~$\kp \ra \kp_c$.
\item \ul{$(S_U + S_\ph)$:} This ``$U\ph$ {\em sector}'' is a typical example of lattice Higgs models, studied intensively nearly 10 years ago (for a review see e.g. ref.~\cc{Je86}). 
Its limit cases at~$\bt \!=\! \infty$ and~$\kp \!=\! 0$ are the pure scalar theory~$S_\ph$ at~$g\!=\!0$, and pure gauge theory~$S_U$, respectively. 
The limit cases~$\bt \!=\! 0$ and~$\kp \!=\! \infty$ have no phase transitions~\cc{OsSe78,FrShe79}, as can be seen by choosing the unitary gauge. 
The magnetic pure scalar field transition point~S extends for~$\bt < \infty$ into a line of Higgs phase transitions, the line~ETS in fig.~\ref{PD},  whereas the confinement-Coulomb transition~C is the basis of the line~CT, T being a triple~point. 
The ETS~line is weakly first order, E being a 2$^{\rm nd}$~order critical endpoint. 
At this point the correlation length, corresponding to the inverse scalar boson mass in lattice units, diverges.

The Higgs mechanism operates above the ETS~line, whereas below the ET~line confinement takes place, and the physical states are gauge invariant composites of~$\ph^\dagger_x$ and~$\ph_y$ or gaugeballs.
The most important are the scalar and vector bosons.
Above the ETS line they correspond to the Higgs and gauge bosons in the gauge invariant formulation of the Higgs mechanism.
Below the ET line they are massive bound states of two massive scalars. 
 
The analytic connection between both regions~\cc{OsSe78,FrShe79} (the so-called {\em complementarity} occuring in the absence of fermions) gave rise to speculations about the strongly coupled standard model~\cc{AbFa81a,AbFa81b,ClFa86}.
However, the transition region to the left of the point~E is difficult to understand in the continuum field theory language (for a recent attempt to explain complementarity see ref.~\cc{Hs93}).
Just this is the region where we expect the \SG\ to operate when the fermions are included. 

There is no local order parameter distinguishing the regions above and below the line~ETS, as~$\phex \!=\! 0$ identically for any~$g > 0$.
Nevertheless, the expectation value of the link product in eq.~(\eq{SPH}), the so-called link energy~$E_{\rm L}$ (see eq.~(\eq{EL})),
is very small below the ETS~line and starts to rise with increasing~$\kp$ above this line, in some reminiscence to the behaviour of~$\phex^2$ in the~$g\!=\!0$ case.
This can be used for a localization of the Higgs phase transition line~\cc{JaJe86,AlAz92,AlAz93}.
Another, more expensive possibility is to calculate the masses (in lattice units) $am_{\rm S}$ and~$am_{\rm V}$ of the scalar and vector bosons. 
They have minima on the ETS~line when considered as  functions of~$\kp$ at fixed~$\bt$~\cc{EvJa87b}. 

We note that the $U\ph$ sector describes the behaviour of all observables constructed from the $U$ and $\ph$ fields in the \chup\ model in the quenched approximation.
This approximation, neglecting virtual fermion loops, is frequently used in numerical simulations of theories with fermion fields on the lattice.

If the constraint $|\ph |=1$ is not imposed, the endpoint E moves to the left as the quartic scalar selfcoupling decreases, eventually reaching $\bt = 0$.

\end{itemize}

\subsection{The relation of the $\chi U \phi$ model to
four fermion theories in the strong coupling limit}
%
At~$\beta \!=\! 0$, the \chup~model in~$d$ dimensions can be rewritten exactly
as a lattice four fermion model~\cc{LeShr87a}. 
In the path integral
\be
   Z = \int \prod_{x, \mu} {\rm d} \chi_x {\rm d} \chb_x
                           {\rm d} \phi_x {\rm d} \phi_x^\dagger
                           {\rm d} U_{x,\mu}
           \exp \{ \,-\, S_{\ch U \ph} \, \}   \; ,
\lb{PI}
\ee
with~$S_{\ch U \ph}$ given in~(\eq{action}), at~$\bt \!=\! 0$ the scalar and gauge fields can be integrated out exactly.
This results in
\be
   Z = r^N J_1^{N_l} \, \int \, \prod_x {\rm d} \chi_x {\rm d} \chb_x
           \exp \{ \,-\, S_{\rm 4f} \, \}    \; ,
\lb{PI4f}
\ee
where~$N$ and~$N_l$ is the number of lattice sites and links, respectively, and
\be
   S_{\rm 4f} = - \sum_x \sum_{\mu = 1}^d \left[ \, G \;  
              \chb_x \chi_x \chb_{x+\mu} \chi_{x+\mu} \, - \,
             {\textstyle \half} \eta_{\mu x}
             \left( \chb_x \chi_{x+\mu} - \chb_{x+\mu} \chi_x \right)
             \right]  \, + \, \frac{am_0}{r} \sum_x \chb_x \chi_x \; ,
\lb{action4f}
\ee
with
\be
\renewcommand{\arraystretch}{1.4}
   r = r(\kp) = \frac{J_U}{J_1}  \;\;\;\;\;\; \mbox{ and } \;\;\;
       \begin{array}{rclll}
         J_U &=& \int {\rm d} U e^{2 \kp {\rm Re} \{ U \} } U &=&
          I_1(2\kp) \;, \\
         J_1 &=& \int {\rm d} U e^{2 \kp {\rm Re} \{ U \} }   &=&
          I_0(2\kp) \;.
       \end{array}
\renewcommand{\arraystretch}{1.}
\lb{rofkappa}
\ee
The fermion field has been rescaled by~$\sqrt{r}$.
The parameter $r$ is an analytic function of~$\kp$ increasing monotonically
from~$r(0) \!=\! 0$ to~$r(\infty) \!=\! 1$.

The action (\eq{action4f}) obviously describes a lattice version of the four-fermion theory. The coupling term contains the nearest neighbour variables because the Grassmann variable $\chi$ has only one component.
The four~fermion coupling parameter~$G$ is related to~$\kp$ via~$r$:
\be
   G := \frac{1-r^2}{4 r^2} \; .
\lb{GNJL}
\ee
From (\eq{rofkappa}) one sees that~$G$ is decreasing monotonically
with increasing~$\kp$;~$G \!=\! \infty$ at~$\kp \!=\! 0$, 
and~$G \!=\! 0$ at~$\kp \!=\! \infty$.
 
As discussed in~\cc{LeShr87a},
in~$S_{\rm 4f}$~(\eq{action4f}) the first and third terms are
explicitly gauge invariant. 
For the second term it should be noted that
the only contributions to the expansion of the exponential
which survive the Grassmann integration are of the form~$\prod_{x \in C} \chb_x \chi_x$, $C$ being closed loops, hence also the
second term is gauge invariant. 

The Lee-Shrock transformation~(\eq{PI}) and~(\eq{PI4f}) is a gauge invariant analogy to the Hub\-bard-Stratonovich transformation used to bilinearize the four fermion actions by introducing an auxiliary scalar field.
The Lee-Shrock transformation introduces for the same purpose both an auxiliary gauge field (there is no pure gauge field term in the action, which corresponds to~$g \!=\! \infty$) and a dynamical scalar field.
The scalar field can be avoided, but the field~$U$ is then a vector auxiliary field without being gauge invariant~\cc{BoKe89}.
The transformation (\ref{PI}) and (\ref{PI4f}) holds only for one staggered fermion field, it can be generalized, however, to the case with several fermion fields.
The resulting pure fermionic theory then contains multifermion couplings of higher degree.
The constraint $|\ph |=1$ is important for the possibility to integrate out the scalar field.

In 4D, the action~(\eq{action4f}) describes the lattice NJL~model~\cc{BoKe89,Ho89,AlGo94}, having \scsb\ and DMG for~$G > G_c \approx 0.28$ corresponding to $\kp < \kp_c \simeq 0.97$.
The character of the phase transition is chiral, i.e. the order parameter is~$\cbcex$.
In lower dimensions the lattice Gross-Neveu models are obtained.

\subsection{Phases of the complete \chup~system in 4D}
%
The phase diagram of the \chup~model with U(1) gauge symmetry in 4D is shown schematically in fig.~\ref{PD}. 
It has been obtained by studying various limiting cases and by numerical simulations with quenched (e.g.~\cc{LeShi86d,LeShr87b,DeShi88,LeShr88b,Shr89,Ku89a}) and unquenched fermions~\cc{DaKo88,AoLe88,Ku89a,DaMe89,MePe91}.

The chiral phase transition of the NJL~model at~$\bt\!=\!0$ (point~N) extends to nonzero values of~$\bt$.
As pointed out in ref.~\cc{LeShr87b}, this can be derived by means of a convergent expansion around $\bt = 0$. 
Therefore, this transition must remain to be of second order at least at small nonzero $\beta$ and cannot end in the interior of the phase diagram.
Furthermore, some properties of the NJL model will persist also at small nonzero $\bt$.

However, the striking observation is that, within the numerical precision, the chiral phase transition joins the Higgs phase transition line~ET at point~E, forming a smooth line~NET. 
This is so both in the quenched and unquenched case, though the fermion feedback changes somewhat the position of the line~ET with respect to the quenched approximation.
The position of the chiral phase transition line~NE changes correspondingly. 
In the vicinity of the point E new critical phenomena, different from the NJL model, might occur.

To our knowledge the interweaving of the chiral and Higgs phase transition is not understood theoretically. 
(When the charges of the scalar and fermion fields are different, it does not occur \cc{LeShi86d,LeShr87b,LeShr88b}.)
Assuming that the transition lines join exactly, there are just two lines of phase transitions, the NETS~line and the CT~line, which separate the~$g \ge 0, \kp \ge 0$ area into three phases: 
\begin{itemize}
\item \ul{Nambu phase.}
This is the area below the NET~line. 
Here the chiral condensate~$\cbcex$ is nonvanishing. 
Both~$\ch$ and~$\ph$ fields are confined, in analogy to the quark confinement in  QCD.
The mass~$am_{\rm F}$ of the fermion state~$F = \pdc$ is nonzero, and thus the DMG occurs.
The justification for the name of this phase is provided by the~$\bt\!=\!0$~limit.
In this limit the $F$ and $\ch$ fermions are identical.
As explained above, here one recovers the four~fermion~model with the mechanism of mass generation, which is traditionally called~``dynamical''.
In this model the ideas of the top quark condensate have been originally formulated \cc{Na89,MiTa89,BaHi90}.

\item \ul{Higgs phase.}
This is the area above the NETS~line. 
Here~$\cbcex\!=\! am_{\rm F} \!=\! 0$ and the Higgs mechanism operates. 
All physical states are gauge invariant and the U(1) charge is screened by the scalar condensate~\cc{EvGr87,EvJa87b}.
Choosing the unitary gauge, one recovers for weak~$g$ by standard perturbative methods the massive U(1) vector boson and the Higgs boson. 
\item \ul{Coulomb phase.}
This is the area below the TS~line. 
The gauge boson is massless and~$\cbcex\!=\! am_{\rm F} \!=\! 0$.
\end{itemize}

A very important observation in 4D is that the line~NE is of second order,  whereas the ETS~line is most probably also in the unquenched case weakly first order~\cc{DaKo88,DaMe89,FrFr95}. 
In this case the point~E is a tricritical~point, where the order of the phase transition changes from the second order (the NE~line) to the weakly first order (the ETS~line).

\section{Dynamical mass generation and chiral phase transition in strongly coupled gauge theories}

Numerous investigations of various strongly coupled gauge theories with fermions on the lattice revealed that the DMG is a generic property of these theories.
It occurs not only in QCD and similar nonabelian theories, but even in the lattice QED at strong coupling, in both its compact~\cc{KoDa87} and noncompact~\cc{KoHa90,GoHo92} formulations.
Can such a naturally occuring dynamical \scsb\ and DMG be of use for the symmetry breaking in the standard model or beyond it, e.g. replacing the Higgs mechanism? 
What is required is a mechanism generating fermion masses, breaking chiral symmetry and providing 3 Goldstone bosons, which is the role of the Higgs-Yukawa sector of the standard model. 
The large top quark mass makes acceptable an approximation to the fermion spectrum, in which only the top quark is massive, the other fermions of the standard model being massless.
In this simplified situation we need to break dynamically a global chiral SU(2) symmetry and ensure that there will be a massive fermion in the spectrum.
This can be the top quark, as suggested in the top condensate models~\cc{Na89,MiTa89,BaHi90,Hi91,Boe91,LiRo92,Ma92}, or possibly some fermion(s) in a new generation.

Except for being nonchiral, the lattice \chup~model with the compact U(1) gauge group is a prototype of the lattice models with such properties.
The DMG occurs naturally at strong coupling and not too large~$\kp$ (the Nambu phase in fig.~\ref{PD}) due to the gauge interaction, i.e. without any help of the charged scalar field~$\ph$. 
The question is how to approach the continuum limit and what are its properties, in particular what spectrum is to be expected.

\subsection{Confinement and spectrum in the Nambu phase}

 In order to find the most suitable way to the continuum, let us first discuss the spectrum within the Nambu phase.
The fundamental fermions are confined, but the scalar field allows us to construct gauge invariant composite fermions of the form~$F = \pdc$, which should correspond to the physical fermions (here we neglect the quark confinement due to QCD).

For an illustration of what can happen in the Nambu phase nonperturbatively, one can think of QCD with a scalar ``quark''~$\varphi$ in addition to the standard quarks~$q$. 
We would then expect the fermionic ``mesons'' of the form~$\varphi^\dagger q$, analogous to the above fermion $F$, the lowest one being stable with respect to the interaction mediated by $A$.
Then there would be many other quark-antiquark~states.
This analogy suggests  the existence  of massive scalar and vector states like~$\ph^{\dagger}\ph$ and~$\cbc$ in the \chup~model.
Thus an occurrence of a Higgs-like bound state is probable. 
The ratio of its mass to the fermion mass will most probably depend on the way the continuum limit is approached.
Possibly  also ``gaugeballs'', states similar to glueballs in the QCD, etc., might exist.
In any case the massless pseudoscalar Goldstone bosons composed of~$\chb$ and~$\ch$ must be present.
On the other hand, the scalar field~$\ph$ itself is confined.

It is important to realize that for small~$\kp$ all states containing~$\ph$ get much heavier than the~$\cbc$~states or gaugeballs, because the constituent~$\ph$ gets infinitely heavy as~$\kp \ra 0$.
As seen from eq.~(\eq{SPH}), the scalar field loses its kinetic term in this limit and simultaneously decouples, as there is no Yukawa coupling. 
At~$g \!=\! 0$, when there is no confinement, one can define the physical  mass~$am_\phi$ of the scalar~$\ph$ as the inverse correlation length in the XY~model, which is known to be infinite at~$\kp \!=\! 0$.
The mass~$am_\phi$ is small only in the vicinity of the point~S and rapidly increases with the distance from this point.
For~$g > 0$ one can think of the ``constituent'' mass of~$\ph$ being roughly equal to $am_\phi$ at the same distance from the line NETS as in the $g = 0$ limit.
That this picture is appropriate has been indicated by the quenched calculations of the masses of the~$\phi^\dagger \phi$~states in the present and similar models~\cc{EvJa87b,EvJe86}.
The masses of various physical states containing~$\ph$ in the \chup\ model can be understood qualitatively by assuming that the constituent mass of~$\ph$ is small along the line~NETS and grows below this line with increasing distance from it, getting infinite at~$\kp \!=\! 0$.
Thus also the fermion mass~$am_F$ in lattice units is large at small~$\kp$ and infinite at~$\kp \!=\! 0$, in agreement with fermion confinement, if the shielding scalar is absent.

\subsection{Where to approach the continuum limit}

As lattice models are applicable to continuum physics only in the vicinity of critical points, where the masses~$am$ in lattice units scale to zero, we have to approach the boundary~NETC of the Nambu phase.
The phase transition along the segment~ET is most probably of first~order  and thus unsuitable.
The phase transition along the CT~line is probably of first~order, too~\cc{KoDa87}.
One can think of changing the gauge part of the action~(\eq{SU}) in such a way that this transition gets continuous~\cc{EvJe85,Ok89}.
However, as discussed above, the scalar constituent of the fermion~$F$ would be heavy at a larger distance from the ETS~line, and the fermion mass~$m_{\rm F}$ would be therefore much larger than that of the various~$\chb \ch$ and gaugeball states.
This is undesirable if we look for an alternative to the Higgs mechanism.

What remains is the NE~phase transition line.
Its most important property is a smooth vanishing of~$\cbcex$
and of~$am_{\rm F}$, like in the NJL model at $\bt = 0$.
However, also other observables, constructed by means of the~$U$
and~$\phi$~fields, can be considered, though they, being of the~$\phi^\dagger \phi$ or gaugeball type, have no natural counterpart in the NJL~model.
The prominent examples are the masses $am_{\rm S}$ and~$am_{\rm V}$ of the scalar and vector bosons occuring already in the spectrum of the $U\phi$ sector.
They are important for the understanding of the Higgs~phase transition at point E and to the right of it.
In particular,~$am_{\rm S}$ vanishes at the point~E. 
Here also the order of the phase transition changes, which in the statistical mechanics is usually associated with a tricritical behaviour (values of critical exponents, etc.)
different from the behaviour at ordinary critical points.
Around the point~E the character of the phase transition changes from
the chiral to the mixed chiral-Higgs phase transition. Therefore, when
considering the approach to the continuum limit we should distinguish
between the vicinity of the point~E and the rest of the NE~line.
For example, quenched calculations suggest that along the NE line, except the point E, no state of the type $\ph^{\dagger}\ph$ scales and thus would not be  present in the continuum spectrum.
Gaugeballs are not expected to scale if the NE line is approached.

It is instructive to elucidate the importance of the scalar field
for the chiral phase transition. 
The DMG at small~$\kp$ is understood as a consequence of the
strong gauge coupling, like in QCD at large distances.
The required fluctuations of the gauge field are, however, gradually
suppressed by the scalar field when~$\kp$ increases.
In the limit~$\kp \ra \infty$, which corresponds to large negative squared bare scalar mass, the gauge field freezes.
This is seen from~(\eq{SPH}) in the unitary gauge,~$\phi_x \!=\! 1$,
when at~$\kp \!=\! \infty$ all the gauge variables are frozen
at the value~$U_{x,\mu} \!=\! 1$, and consequently the chiral condensate is zero.
So the chiral transition on the line~NET takes place when the 
suppression of the gauge field fluctuations by the scalar field is
sufficiently strong.
This is consistent with the picture in the~NJL~model
at~$\bt\!=\!0$:
here the DMG ends when the four~fermion coupling,
which according to~(\eq{GNJL}) decreases with increasing~$\kp$,
is sufficiently small.

\subsection{The shielded gauge mechanism in continuum}

We are now in a position to formulate the idea of the SGM in the \chup\ model.
Essentially it is a possibility that this model, or some suitable generalization of it, is renormalizable in the vicinity of the NE line and thus can be used for the continuum physics in a large range of energies.
This is meant in the following nonperturbative sense:

On the lattice the concept of renormalizability is conveniently formulated in terms of the lines of constant physics.
These are the lines in the bare parameter space along which the dimensionless ratios of physical observables stay approximately constant as the cutoff is varied.
Such lines, if at all, are found in the vicinity of critical manifolds in their scaling regions, where the lattice artefacts are negligible.
Lines of constant physics can hit the critical manifold, i.e.\ the masses~$am$ in lattice units get zero, as they are inverse correlation lengths.
Then these lines can be used for continuum physics without any intrinsic energy restriction, because the lattice cutoff can be completely removed, as for example in the asymptotically free theories.
It can also happen that lines of constant physics approach the critical manifold, yet remain in some small but finite distance from it.
The cutoff can then be made large, but has to be kept finite, and the lines can be used for continuum physics only in some energy range substantially smaller than the cutoff.
This is the situation in the trivial theories like the Higgs-Yukawa sector of the standard model.
Both types of theories can be considered as renormalizable, in the second case with an energy restriction.

For the SGM to operate it is necessary that the \chup\ model with the U(1) gauge symmetry, or some similar model, have lines of constant physics in the Nambu phase in the vicinity of the NE~line.
The NJL~model~points at~$\bt \!=\! 0$ can be special points of these lines, but we cannot exclude that these lines completely avoid the~$\bt \!=\! 0$ line in fig.~\ref{PD}.
We do not insist that the cutoff can be removed completely, i.e. that a nontrivial fixed point exists.
This would be nice and is not impossible but, as the success of the standard model shows, not necessary.

It can happen that in a given model the lines of constant physics do not exist.
This is apparently the case in the lattice NJL~model~\cc{AlGo94}.
In such a situation one may try to generalize the model by introducing new fields and/or couplings, so that the parameter space is enlarged by a few relevant parameters, and the lines of constant physics can be found.
One well known example is the nonlinear~$\sg$ model in 4D with its generalization to the full~$\phi^4$ theory.
Also the NJL~model can be generalized to the Yukawa model by changing the auxiliary character of the scalar field into a dynamical one~\cc{HaHa91,Zi91,BoDe92b,DeJe92}.
The strongly coupled QED on the lattice is another example of a model in which the existence of the lines of constant physics is at least disputed~\cc{KoHa90,GoHo92}, but its appropriate generalization has not yet been found.

We have two qualitative arguments for the expectation that the \SG\ could operate in the \chup\ model as defined in sec. 2.1.
First, the generalization from the NJL~model to the \chup\ model is similar to the step from the auxiliary to the dynamical scalar field generalizing successfully the NJL to the Yukawa model.
Second, as~$\bt$ increases, the character of the NE~line gets more complicated than in the NJL case.
As we have discussed above, the chiral character gets mixed with the Higgs type of phase transition, in particular around the point~E.
This means that new relevant parameter(s) may come into game.
This should be expected in particular if the point~E is a tricritical~point. 

However, it could be necessary to generalize the model by adding one or two more couplings, possibly in a similar way as the four fermion theories are generalized to Yukawa theories \cc{HaHa91,Zi91,BoDe92b}.
To explain this generalization we first recall that the four fermion theories are usually bilinearized by introducing an scalar auxiliary field, which we call~$\Om$.
Its symmetry properties are those of the global chiral symmetry group.
Such a transformation of a four fermion theory opens space for a generalization by including also the kinetic and selfcoupling term of the~$\Om$ field into the action, obtaining in this way a Yukawa model~\cc{HaHa91,Zi91,BoDe92b}.
In 4D this generalization makes from the NJL~model a renormalizable theory.
For a strong Yukawa coupling the field~$\Om$ can be understood as a composite of~$\cbc$.
An analogous generalization of the \chup\ model might thus consist in introducing a composite scalar field~$\Om = \cbc$ with charge zero and including into the action (\ref{action})  also the kinetic and selfcoupling terms of this field. 
One might hope that the model has similar renormalizability properties as strongly coupled Yukawa models.
These models have been found to have DMG and fit well into scenarios with \SG. 
We note that some gauged Yukawa models have been considered \cc{Ho91,Ko93,Kr93,KoShi94} for similar purposes.
Finally, it could be also important to relax the constraint $|\ph |$=1.
Unfortunately, with such amendments the models with SGM would get quite complex.

The idea of the \SG\ is profoundly nonperturbative and requires numerical verification in a very difficult regime of the lattice field theories around the point E: the gauge fields are strongly coupled, i.e. strongly fluctuating, and ultimately the SGM should be investigated with dynamical fermions.
However, as in QCD, it could be that many essential features of the \chup\ models, in particular the DMG, can be studied in the quenched approximation, which is the aim of the next section.

\section{Some numerical results in the quenched approximation}

Our nonperturbative numerical results in the strong coupling region, obtained up to now, confirm some of the anticipated features of the Nambu phase, which are necessary for the \SG\ to work:
In the phase with \scsb, where~$\cbcex$ is nonzero, also the fermion~mass in lattice units~$a m_{\rm F}$ is nonzero, implying DMG.
Both observables seem to scale to zero when the transition line~NE is approached. This is so even in the vicinity of the point E, whereas along the ET line the transition is apparently of first order also in fermionic observables.
The results for~$\cbcex$ and~$a m_{\rm F}$ are also consistent with
the prediction of the gap~equation relating these two quantities and obtained from the spectral function of the free fermion with the physical mass $am_F$.
The pseudo-scalar fermion-antifermion state $\pi$ behaves in the Nambu phase like a Goldstone~boson, i.e.~$(a m_\pi)^2$ is a linear function of the bare fermion~mass~$a m_0$. 
This behaviour is unique for the Nambu phase.

We have initiated the study of further states in the spectrum
which we will use in our forthcoming work \cc{FrFr95} in the search for the lines of constant mass ratios.
The mass of the scalar fermion-antifermion state $\sg$ has been determined only very roughly, however. 
This state might be a candidate for the effective (composite) Higgs boson.
At the moment we can only say that the staggered fermions can be used in this kind of models as at least in the pseudoscalar and vector channels no problems with the flavour symmetry restoration have been detected.
These results are described in some detail in this rather technical section.

\subsection{Definitions of the observables}
%
For the localization of the phase transition lines and the
determination of the particle spectrum the following
observables were used: 

The normalized plaquette and link energies, defined as
\bea
    E_{\rm P} &=& \frac{1}{6V} \sum_{\rm P} 
                    {\rm Re} \{ U_{\rm P} \} \;, \lb{EP} \\
    E_{\rm L} &=& \frac{1}{4V} \sum_{x,\mu} 
                    {\rm Re} \{ \phi^\dagger_x U_{x,\mu} \phi_{x+\mu} \} \;,
\lb{EL} 
\eea
where~$V\!=\!L^3 T$ is the lattice volume.
These observables have been used for the localization of 
the Higgs phase transition in the $U\ph$ sector
(see section~\ref{subsystems})~\cc{JaJe86,AlAz92,AlAz93},
i.e.\ in the quenched approximation of the \chup~model.
In particular, in ref.~\cc{AlAz93}~$E_{\rm P}$ and~$E_{\rm L}$
where used in the framework of the multihistogram method
for a high-precision determination of the position of
the endpoint~E in this approximation.

For the localization of the chiral phase transition line,
in particular the segment~NE,
we must use fermionic observables. In the quenched approximation
this line can only be seen in fermionic quantities because 
of the missing feedback of the fermions to the bosonic fields.
We measure the chiral~condensate with the stochastic~estimator~method:
\be
   \cbcex = \left\lag \frac{1}{V} {\rm Tr} \{ M^{-1} \} \right\rag
      \approx \left\lag \vec{\eta}^{\,\dagger} M^{-1} \vec{\eta} \right\rag \;,
\lb{cbc}
\ee
where~$\vec{\eta}$ is a vector of dimension~$V$ filled with 
gaussian random numbers (see,~e.g.,~\cc{BiKe89}).
It should be noted that with fixed-length scalar field~$|\phi_x| \!=\! 1$
this condensate~$\cbcex$ coincides with the 
condensate~$\lag \overline{F} F \rag$ constructed from the
neutral fermionic fields~$F$ defined below.

The next important fermionic quantity is the mass of the physical fermion.
We expect in the spectrum of the Nambu phase a neutral fermionic state whose mass~$a m_{\rm F}$ is nonzero and scaling to zero when the chiral~transition line~NE is approached.
We have considered the gauge invariant fermionic field
\be
   F_x := \phi^\dagger_x \chi_x \;, \hspace*{2cm} 
   \overline{F}_x = \phi_x \chb_x \; , 
\lb{FFbar}
\ee
and determined numerically the corresponding fermion propagator
\be
\renewcommand{\arraystretch}{0.5}
   G_{\rm F}(t) = \frac{2^3}{V} \sum_{\begin{array}{c} \scriptstyle \vec{x} \\*[0.5mm]
                              \scriptstyle x_1,x_2,x_3 \\ 
                              \scriptstyle {\rm even}
                              \end{array}}
                        \sum_{\begin{array}{c} \scriptstyle \vec{y} \\*[0.5mm]
                              \scriptstyle y_1,y_2,y_3 \\ 
                              \scriptstyle {\rm even}
                              \end{array}}
           \left\lag F_{\vec{x},t} \overline{F}_{\vec{y},0} 
           \right\rag \;.
\lb{GF}
\renewcommand{\arraystretch}{1}
\ee
The sums mean that we use a special wall source
where only one point per hypercube is set to~1, the remaining ones to~0,
as usual for staggered fermions (see e.g. \cc{GuGu91}).
The numerical data for~$G_{\rm F}(t)$ 
are fitted with MINUIT to the Ansatz
\be
   G_{\rm F}(t) = A_{\rm F}
          \left( e^{- E_{\rm F} t} - (-1)^t e^{- E_{\rm F}(T-t)} 
          \right) \;.
\lb{GFfit}
\ee
The mass $a m_{\rm F}$ of the gauge invariant fermion and the corresponding wave function renormalization constant~$Z_{\rm F}$
are then 
\bea
    a m_{\rm F} &=& \sinh E_{\rm F} \;,  \label{mFfromEF} \\
      Z_{\rm F} &=& A_{\rm F} \left( 1 + e^{- E_{\rm F} T} \right) \cosh E_{\rm F} \;.   \label{ZFfromAFEF} 
\eea

Having in mind our long-term aim to determine lines of constant 
physics, i.e.\ lines of constant mass ratios, we have started to look
at further states in the spectrum. 
First we define the fermion-antifermion composite states,
the ``mesons''.
Table~\ref{tabmesons} shows a list of the operators in the
more familiar continuum notation and the corresponding
translation to the formulation with staggered fermions. 
Lattice experts will notice that we use the four simplest local operators from the Golterman~tables~\cc{Go86}.
We note that it is irrelevant whether these operators are
constructed with the $\chi$~field or with the neutral~$F$~field~(\eq{FFbar})
because of the fixed length of the~$\phi$~field.

\begin{table}
\renewcommand{\arraystretch}{1.5}
\begin{center}
\begin{tabular}{|r|c|r|c|l|l|} \hline
\vphantom{$\ds \sum^a$} $i$ & continuum & staggered fermions & $s^{ik}_x$ & $J^{PC}$ & particle \\*[1mm] \hline\hline
1&$\psb \psi$&$\chb_x \chi_x$&1&$\begin{array}{l}
                               0^{++}_s \\ 0^{-+}_a
                             \end{array}$ & 
                            $\begin{array}{l}
                               \sigma \; {\rm (f}_0) \\ \pi^{(1)}
                             \end{array}$ \\ \hline
2&$\psb \gm_5 \psi$&$\eta_{4x}\xi_{4x} \chb_x \chi_x$&$\eta_{4x} \xi_{4x}$&
                            $\begin{array}{l}
                               0^{+-}_a \\ 0^{-+}_a
                             \end{array}$ & 
                            $\begin{array}{l}
                               -  \\ \pi^{(2)}
                             \end{array}$ \\ \hline
3&$\psb \gm_k \psi$&$\eta_{kx} \vep_x \xi_{kx} \chb_x \chi_x$&$\eta_{kx} \vep_x \xi_{kx}$&
                            $\begin{array}{l}
                               1^{++}_a \\ 1^{--}_a
                             \end{array}$ & 
                            $\begin{array}{l}
                               a \\ \rho^{(3)}
                             \end{array}$ \\ \hline
4&$\psb \gm_k \gm_5 \psi$&$\eta_{4x} \xi_{4x} \eta_{kx} \vep_x \xi_{kx} \chb_x \chi_x$&$\eta_{4x} \xi_{4x} \eta_{kx} \vep_x \xi_{kx}$&
                            $\begin{array}{l}
                               1^{+-}_a \\ 1^{--}_a
                             \end{array}$ & 
                            $\begin{array}{l}
                               b \\ \rho^{(4)}
                             \end{array}$ \\ \hline\hline
\end{tabular}
\renewcommand{\arraystretch}{1}
\caption%
[Mesonic operators]%
{
List of mesonic operators in familiar continuum notation
and in the staggered fermion formulation, together with
the continuum quantum number assignement~$J^{PC}$ and the names
of the corresponding QCD particles. 
The sign factors~$s^{ik}_x$ are composed of the standard
staggered phase factors~$\eta_{\mu x} = (-1)^{x_1+ \cdots +x_{\mu-1}}$,
$\xi_{\mu x} = (-1)^{x_{\mu+1}+ \cdots +x_4}$ 
and~$\vep_x = (-1)^{x_1 + \cdots + x_4}$.
}
\label{tabmesons}
\end{center}
\end{table}
The timeslice operators for the mesons are given by
\be
   {\cal O}^{ik} (t) = \sum_{\vec{x}} s^{ik}_{\vec{x},t} \chb_{\vec{x},t}
                        \chi_{\vec{x},t} \;,
\lb{Omesons}
\ee
with~$s^{ik}_x$ given in table~\ref{tabmesons}.
We measure the correlation functions
\be
   G^{(i)}_{\rm mesons} (t) = \frac{1}{N^{(i)}_k} \sum_k^{N^{(i)}_k}
          \left\lag {\cal O}^{ik} (t) {\cal O}^{ik} (0) 
                 \right\rag \;, \;\;\; i = 1,\ldots,4 \;,
\lb{Gmesons}
\ee
where~$N_k^{(i)} \!=\! 1$ for~$i \!=\! 1$,~2 and~$N_k^{(i)} \!=\! 3$
for~$i \!=\! 3$,~4, i.e.\
for~$i\!=\!3$,~4 we average
over the space directions~$k\!=\!1,\ldots,3$.
For the actual measurement~$G^{(i)}_{\rm mesons}$ has to be expressed
in terms of the inverse fermion matrix~$M^{-1}$:
\bea
   G^{(i)}_{\rm mesons} (t) &=& \frac{1}{N_k^{(i)}} \sum_k^{N_k^{(i)}}
                  \left\lag \sum_{\vec{x} \vec{y}}
                  s^{ik}_{\vec{x},t} \chb_{\vec{x},t} \chi_{\vec{x},t}
                  s^{ik}_{\vec{y},0} \chb_{\vec{y},0} \chi_{\vec{y},0}
                 \right\rag  \nonumber \\
               &=& - \frac{1}{N_k^{(i)}} \sum_k^{N_k^{(i)}}
                     \left\lag \sum_{\vec{x} \vec{y}}
                  s^{ik}_{\vec{x}} s^{ik}_{\vec{y}}
                  (M^{-1})_{(\vec{x},t),(\vec{y},0)}
                  (M^{-1})_{(\vec{y},0),(\vec{x},t)}
                 \right\rag \;
\lb{Cm2}
\eea
up to a constant disconnected term.
In the second line we used the fact that the~$s^{ik}$ factorize according to:
\be
   s^{ik}_x = s^{ik}_{\vec{x},t} = s^{ik}_t s^{ik}_{\vec{x}} =
               s^{ik}_{\vec{x}} \;,
\lb{sik}
\ee
because~$s^{ik}_t \!=\! 1$ for all~$(ik)$.
The relation
\be
  (M^{-1})_{(\vec{x},t),(\vec{y},0)}
  (M^{-1})_{(\vec{y},0),(\vec{x},t)} = 
   \vep_{\vec{y},0} \vep_{\vec{x},t}
  (M^{-1})_{(\vec{x},t),(\vec{y},0)}
  \left[ (M^{-1})_{(\vec{x},t),(\vec{y},0)} \right]^*  
\lb{MMeeMM}
\ee
shows that actually only one fermion matrix inversion
is needed per source~point.
With
\be
   \vep_{\vec{x},t} = (-1)^{t} \, \eta_{4 \vec{x}}
\ee
we finally have
\be
   G^{(i)}_{\rm mesons} = - (-1)^t \frac{1}{N_k^{(i)}} \sum_k^{N_k^{(i)}}
            \left\lag \sum_{\vec{x},\vec{y}}
               s^{ik}_{\vec{x}} s^{ik}_{\vec{y}}
               \eta_{4 \vec{x}} \eta_{4 \vec{y}}
               (M^{-1})_{(\vec{x},t)(\vec{y},0)}
            \left[ (M^{-1})_{(\vec{x},t)(\vec{y},0)}
            \right]^* \right\rag \;.
\ee
In our simulations we use point~sources for this measurement.

We fit these~$G^{(i)}_{\rm mesons}$ to the Ansatz~\cc{AlGo94}
\bea
   G_{\rm mesons} (t) &=& \phantom{(-1)^t} \sum_n A^+_n 
         \left( e^{-E^+_n t} + e^{-E^+_n (T-t)} \right) \nonumber \\
               &+&          (-1)^t  \sum_n A^-_n 
         \left( e^{-E^-_n t} + e^{-E^-_n (T-t)} \right) \nonumber \\
               &+& \phantom{(-1)^t} B^+ \nonumber \\
               &+&          (-1)^t  B^-  \;.
\lb{Gmesonsfit}
\eea
Here $E^+$ is the energy of the s-wave states whereas $E^-$ that of the parity  partners, which are p-waves.
The masses of the mesons are then computed from these energies by

\be
   a m = 2 \sinh ( E / 2) \;.
\lb{mfromE}
\ee

Note that already this minimal set of~$4$~mesonic operators
given in table~\ref{tabmesons} allows us to check
for flavour symmetry restoration, because there are
two states having overlap with
two different operators: the~$\pi$ shows up in~${\cal O}^{(1),(2)}$
and the~$\rho$ in~${\cal O}^{(3),(4)}$.

We are further interested in the scalar and vector bosons, which
are present also in the~$U \phi$~sector without fermions.
The corresponding operators are
\bea
    {\cal O}^{({\rm S})} (t) &=& \frac{1}{L^3} \sum_{\vec{x}} {\rm Re} 
              \left\{ 
       \sum_{i=1}^{3} \phi^\dagger_{\vec{x},t} U_{(\vec{x},t), i} 
                      \phi_{\vec{x}+\vec{i},t} \right\} \;,
\lb{OS} \\
    {\cal O}^{({\rm V})}_{i} (t) &=& \frac{1}{L^3} \sum_{\vec{x}} {\rm Im} 
              \left\{ \phi^\dagger_{\vec{x},t} U_{(\vec{x},t), i} 
                      \phi_{\vec{x}+\vec{i},t} \right\} \;, \;\;\;\;\;\;
                             i = 1,\;2,\;3 \;,
\lb{OV}
\eea
and the correlation functions are
\bea
    G^{({\rm S})} (t) &=& \left\lag {\cal O}^{({\rm S})} (t)
                                    {\cal O}^{({\rm S})} (0) 
                          \right\rag \;, \lb{GS} \\
    G^{({\rm V})} (t) &=& \frac{1}{3} \sum_{i=1}^3 \left\lag 
                                    {\cal O}^{({\rm V})}_i (t)
                                    {\cal O}^{({\rm V})}_i (0) 
                          \right\rag \;. \lb{GV} 
\eea
We fit them to a simple~$\cosh$-Ansatz:
\be
   G (t) = A \left( e^{-E t} + e^{-E (T-t)} \right) \;.
\lb{GSVfit}
\ee
The masses of the scalar boson~$a m_{\rm S}$ and the vector boson~$a m_{\rm V}$
are then computed from the energies~$E_{\rm S,V}$ by means of the formula~(\eq{mfromE}).

\subsection{Numerical investigation of the NE~line 
            in the quenched approximation}
\begin{figure}
\centerline{                                                                    
\fpsxsize=9.0cm                                                                 
\fpsbox{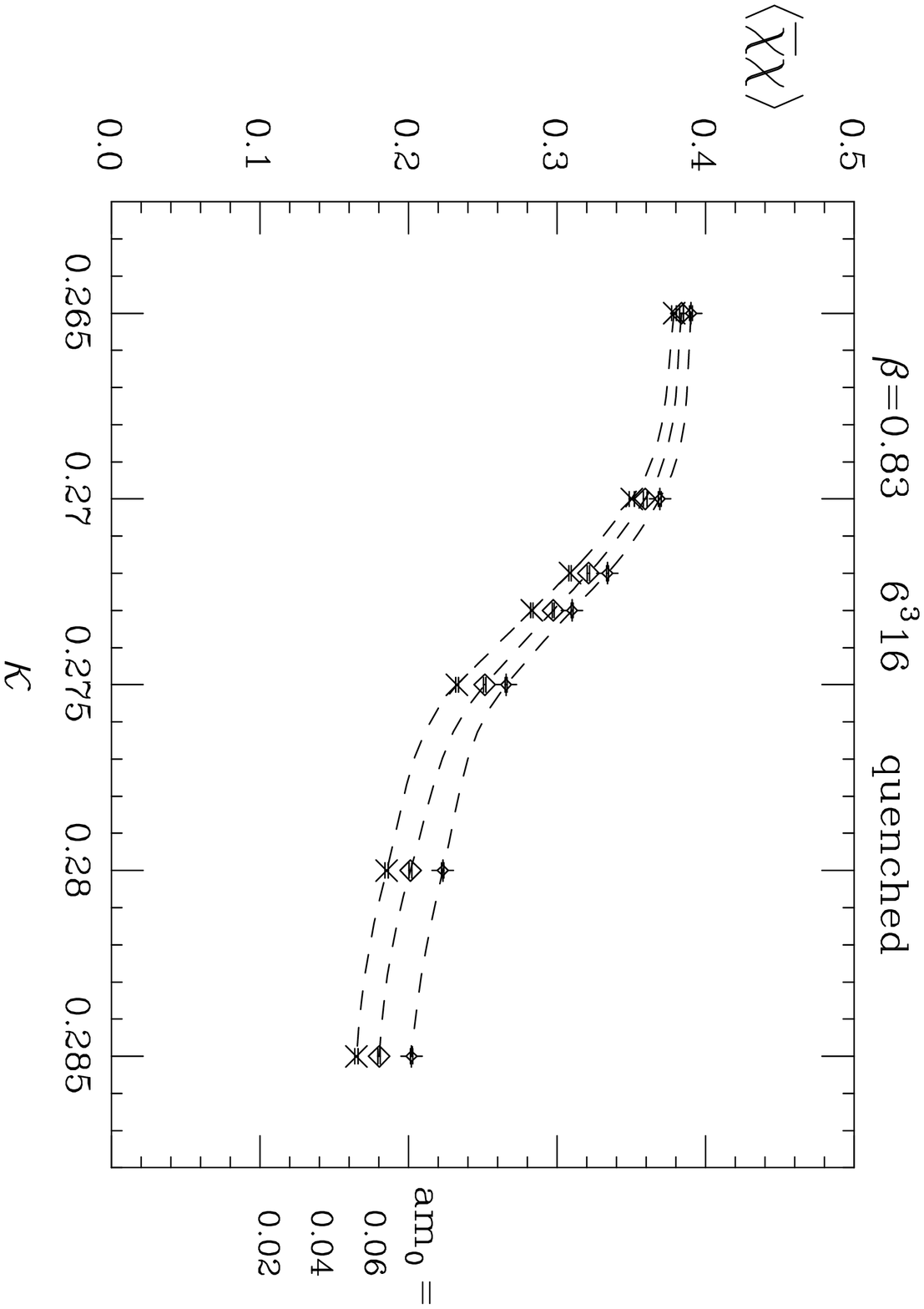}                                                               
}                                                                               
\centerline{                                                                    
\fpsxsize=9.0cm                                                                 
\fpsbox{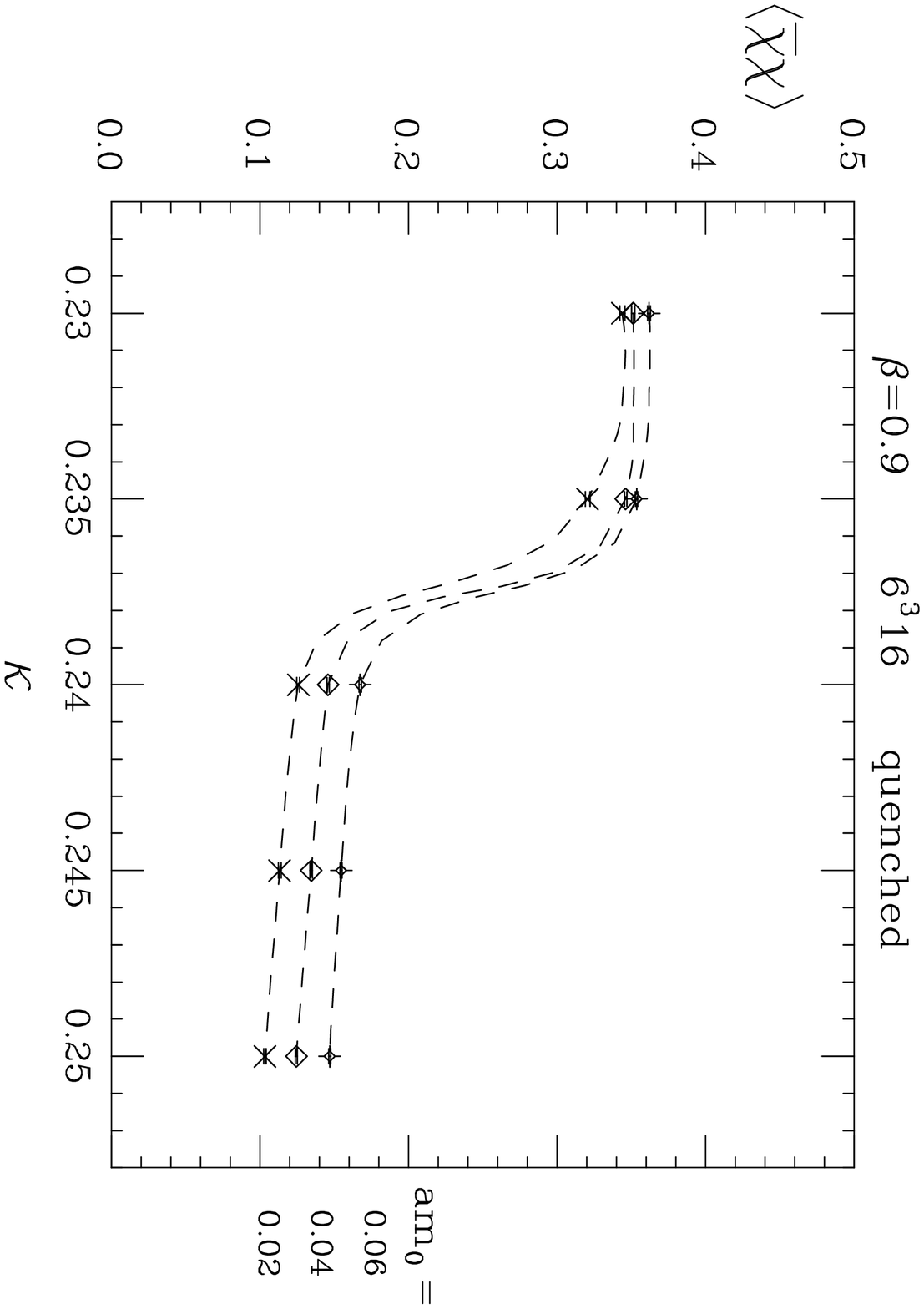}                                                               
}                                                                               
\caption%
[...caption in list of figures...]%
{ 
The chiral condensate as a function of~$\kp$ at two~$\bt$~values:
one slightly below~($\bt = 0.83$, upper figure) and one slightly
above~$\bt_{\rm E} \approx 0.85$~($\bt = 0.90$, lower figure).
The dashed lines are included to guide the eye.
}
\label{figcbc}                                                                   
\end{figure}

\begin{figure}
\centerline{                                                                    
\fpsxsize=9.0cm                                                                 
\fpsbox{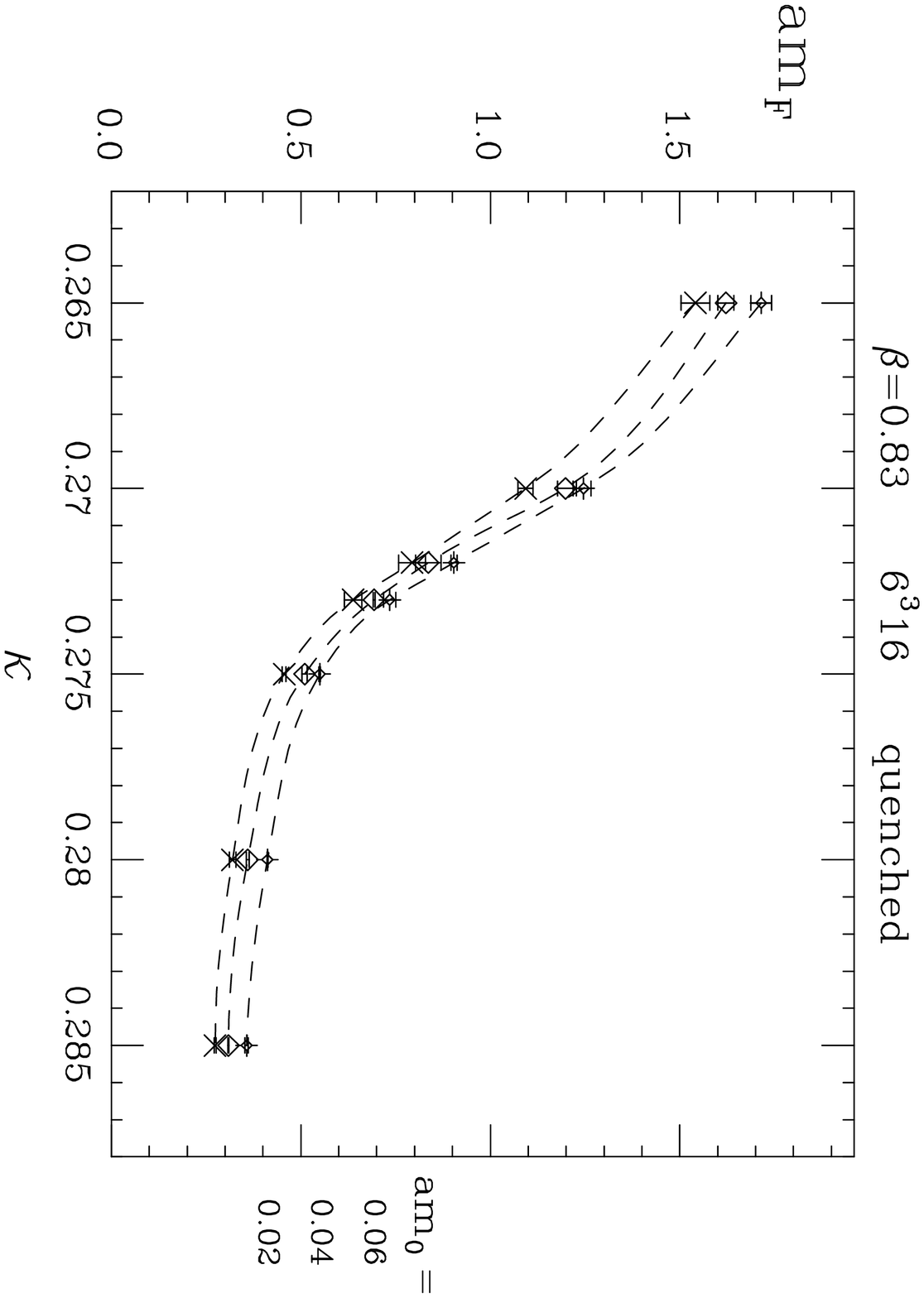}                                                               
}                                                                               
\centerline{                                                                    
\fpsxsize=9.0cm                                                                 
\fpsbox{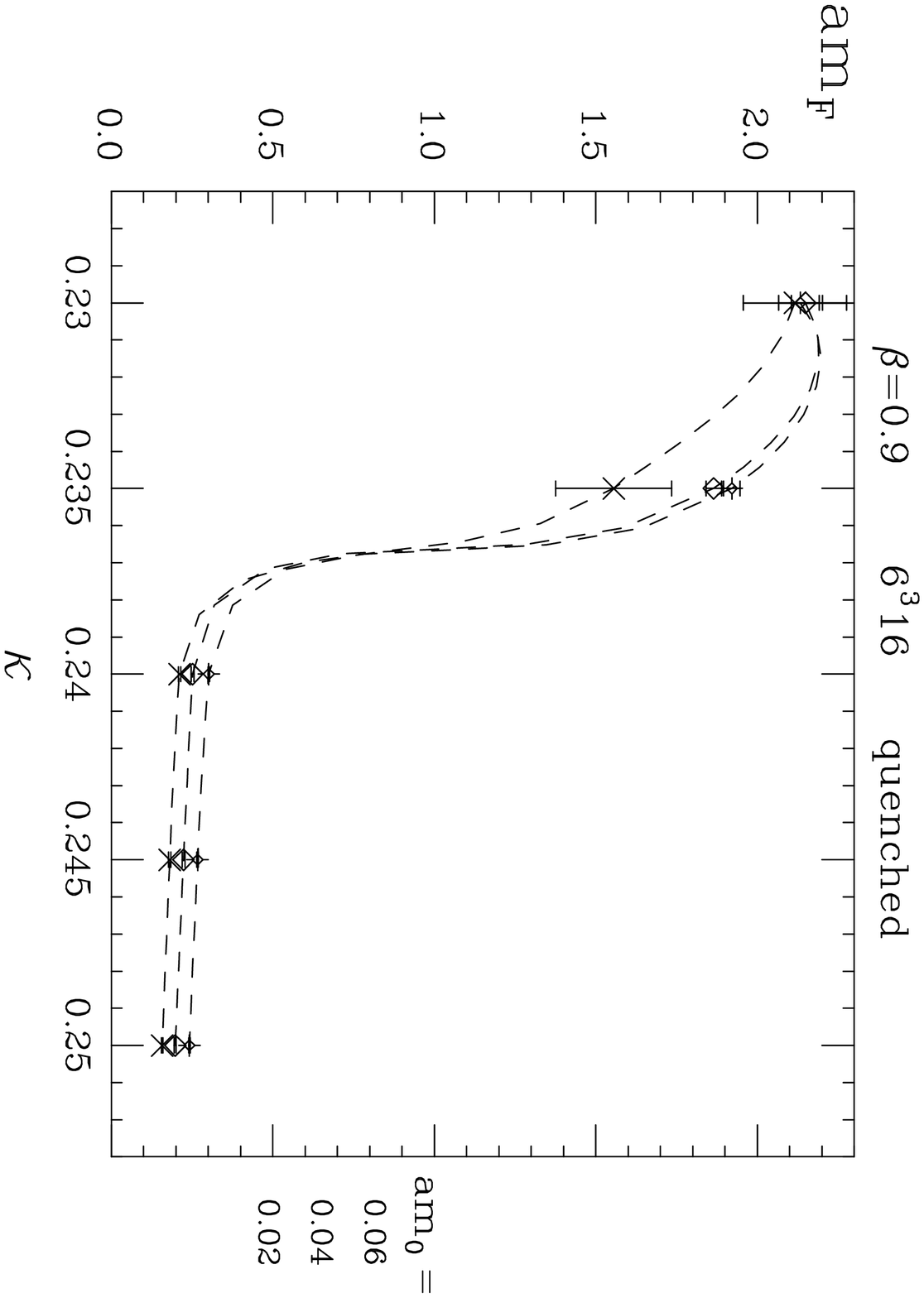}                                                               
}                                                                               
\caption%
[...caption in list of figures...]%
{ 
The fermion mass~$a m_{\rm F}$ as a function of~$\kp$ at two~$\bt$~values:
one slightly below ($\bt = 0.83$, upper figure) and one slightly
above~$\bt_{\rm E} \approx 0.85$~($\bt = 0.90$, lower figure).
The dashed lines are included to guide the eye.
}
\label{figmF}                                                                   
\end{figure}

%
As already stated above, the existing lattice results for the
magnetic transitions in the~$U \phi$~sector~\cc{JaJe86,AlAz92,AlAz93}
are also relevant for the \chup~model in the quenched approximation,
and thus the only new feature of the phase diagram with quenched fermions
is the chiral~transition.
The Lee-Shrock transformation~(\eq{PI}),~(\eq{PI4f}) in combination
with mean field theory predicts a
critical point at~$\bt\!=\!0$ and~$\kp_c \approx 1.15$ 
for~$N_f \!=\! 4$,~\cc{LeShr87a} with
\scsb\ below~$\kp_c$ and chiral~symmetry restauration above~$\kp_c$.
The first question is whether this chiral phase transition
continues to finite~$\bt$ and where it runs to.
Earlier quenched investigations~\cc{LeShi86d,LeShr87b,DeShi88} already come to the result
that the chiral~phase~transition joins the line~ET.
Our comprehensive results for~$\cbcex$ and~$a m_{\rm F}$ confirm
this observation. 
In particular on the line~ET, where the bosonic observables indicate
a~$1^{\rm st}$~order transition, we also observe jumps in~$\cbcex$
at the same places.
On the other hand, along the line~NE, even quite close to point~E,
we do not observe any indication for a discontinuous phase transition.
This is demonstrated 
in figure~\ref{figcbc}, where we compare the numerical results for~$\cbcex$
as a function of~$\kp$ at~$\bt$~values slightly below and 
above~$\bt_{\rm E} \!=\! 0.8485(8)$~\cc{AlAz93}.
%
%
The fermion~mass~$a m_{\rm F}$ shows a similar behaviour, as is shown
in figure~\ref{figmF}.
From these numerical results we conclude that in the quenched approximation
the chiral phase transition line joins the line~ET, as shown in
the schematic phase diagram in figure~\ref{PD},
changing at the point~E
from~$2^{\rm nd}$~order on the~NE~line to~$1^{\rm st}$~order on the~ET~line.

%
%
                                                                              
Figure \ref{figmF} also demonstrates the DMG below the line~NE, where the fermion mass is clearly nonzero and seems to scale to zero when this line is approached.
The finite~$a m_{\rm F}$ values in the chirally symmetric phase are probably
an artefact of the small lattice volume and the finite bare~mass~$m_0$,
but of course this conjecture has to be confirmed by some theoretically
well-funded extrapolation scheme to the thermodynamic limit and
to~$m_0 \ra 0$, e.g. by methods developed in refs.~\cc{LeShr87b,LeShr88a,Ho89}. 

\begin{figure}
\centerline{                                                                    
\fpsxsize=12.0cm                                                                 
\fpsbox{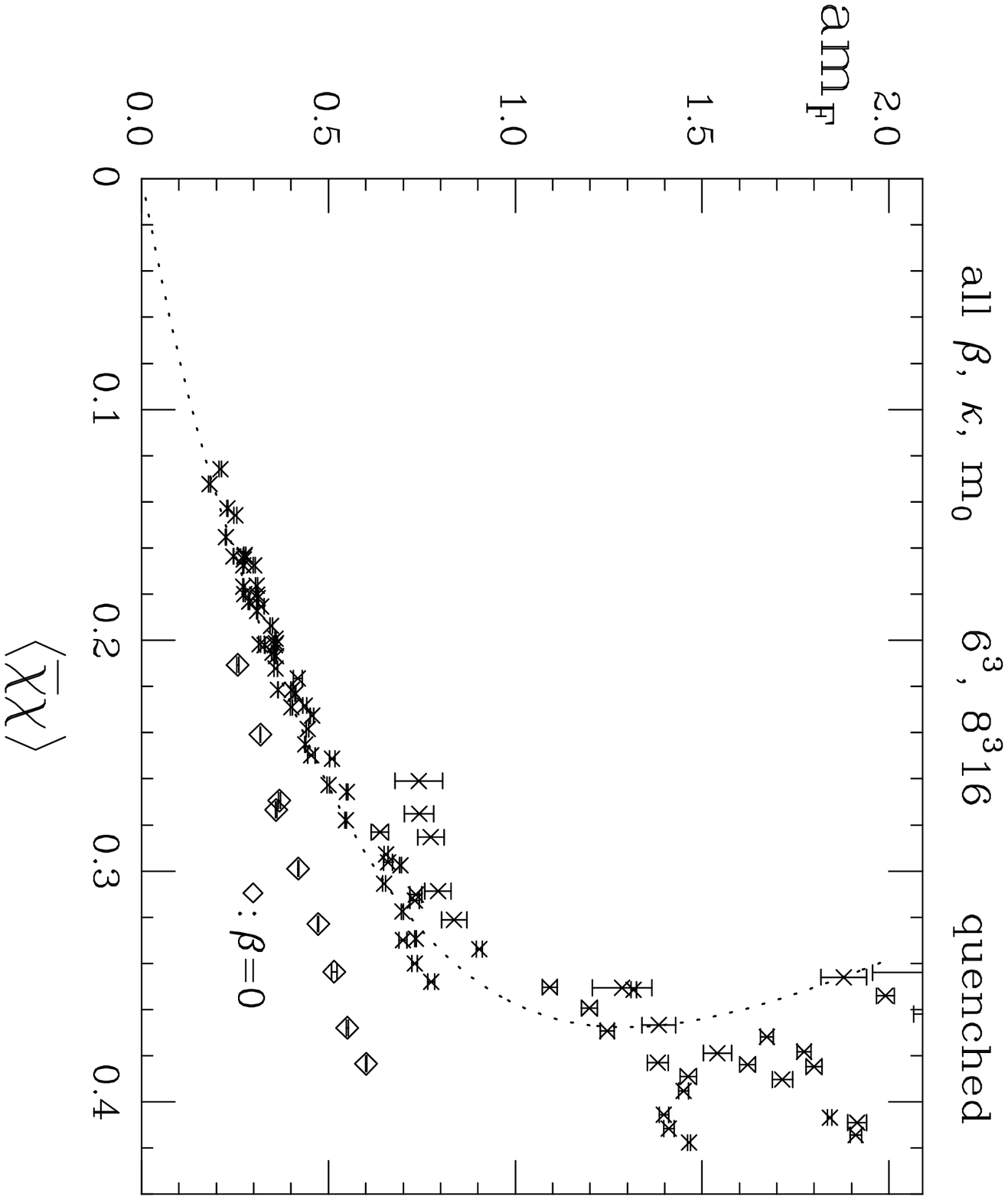}                                                               
}                                                                               
\caption%
[...caption in list of figures...]%
{ 
All our quenched results for~$a m_{\rm F}$ as a function of~$\cbcex$.
The dotted line is the prediction of the gap~equation~({\protect\eq{gapeq}}).
}
\label{figmFcbc}                                                                   
\end{figure}

In figure~\ref{figmFcbc} we plot our numerical results for~$a m_{\rm F}$ 
as a function of~$\cbcex$ together with the relation between them, which follows from the free fermion propagator with the observed fermion mass (the gap equation):  
\be
   \cbcex = \frac{1}{V} \sum_{p_\mu} 
       \frac{a m_{\rm F}}{(a m_{\rm F})^2 + \sum_\mu \sin^2 p_\mu } \;.
\lb{gapeq}
\ee 
Here the sum runs over the set of momenta corresponding to
a finite lattice with periodic (antiperiodic) boundary~conditions 
in the space (time) directions, as we have it in our simulations.
Nearly all the data (except at~$\bt\!=\!0$) lie in a narrow band
around the curve (\ref{gapeq}), indicating that the fermion wave function renormalization constant is close to one.
Such a behaviour was previously observed also in
strongly coupled noncompact~QED on the lattice~\cc{GoHo92} 
and in the NJL~model~\cc{AlGo94}, both with dynamical fermions.
%
%
 
\begin{figure}
\centerline{                                                                    
\fpsxsize=12.0cm                                                                 
\fpsbox{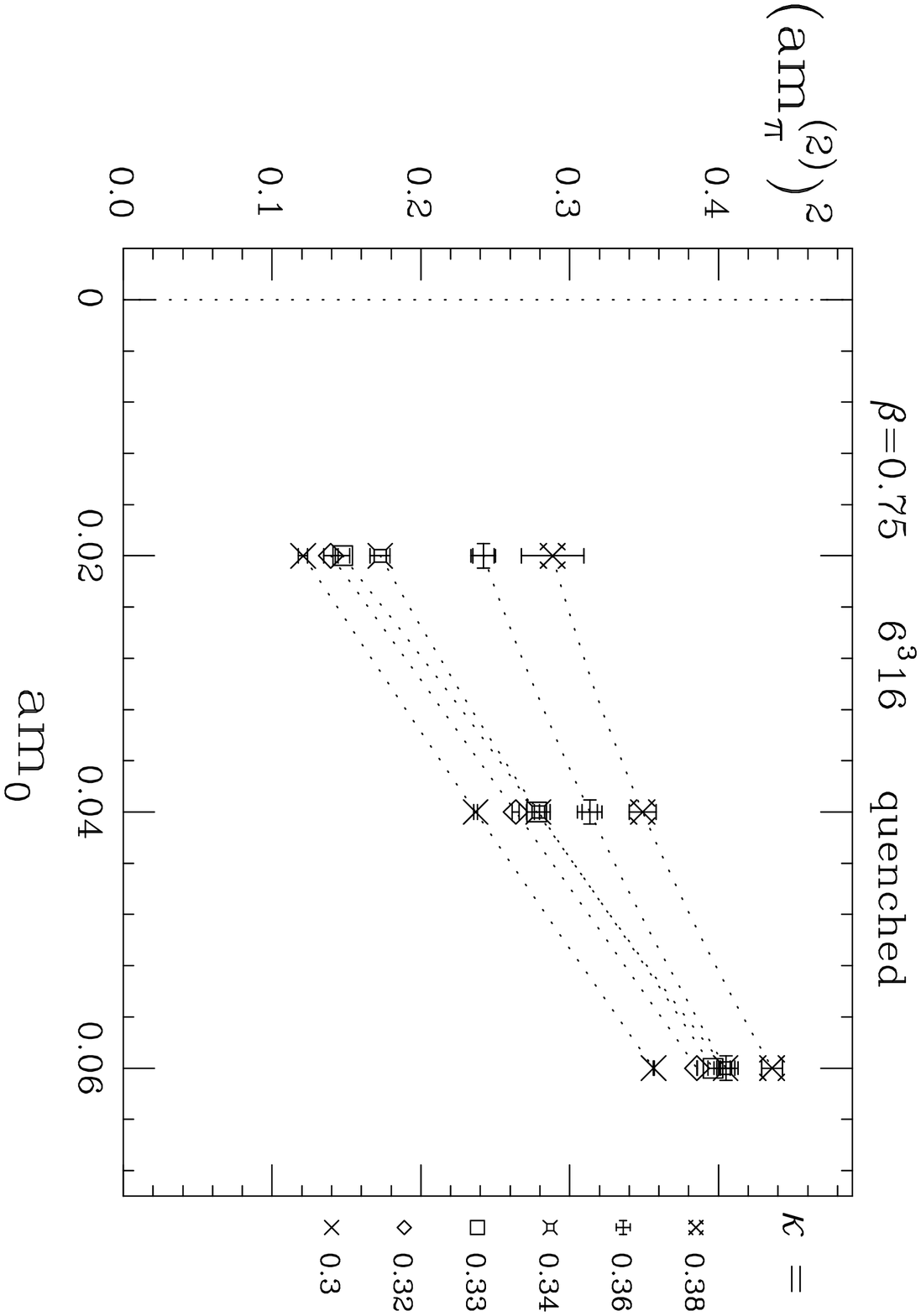}                                                               
}                                                                               
\caption%
[...caption in list of figures...]%
{ 
The square of the pseudoscalar~mass~$(a m_\pi)^2$ as a function of~$a m_0$
(it is denoted~$a m_\pi^{(2)}$ because it was measured with the second operator in table~{\protect\ref{tabmesons}};~$a m_\pi^{(1)}$, measured with the first operator, is consistent with~$a m_\pi^{(2)}$ but is much more noisy).
For~$\kp$ within the Nambu phase~$(a m_\pi)^2$ is linear in~$a m_0$,
for the higher~$\kp$ deviations are observed.
}
\label{figmpi2}                                                                   
\end{figure}

An indicator for \scsb\ is the 
pseudo-scalar~mass~$a m_\pi$ as a function of~$a m_0$:
in the chirally broken phase the ``$\pi$ meson'' should behave
like a Goldstone~boson, i.e.~$(a m_\pi)^2$ should go to zero linearly 
as a function of~$a m_0$. 
Conversely, one should see deviations 
from such a behaviour in the chirally symmetric phase.
In our numerical results for~$a m_\pi$ we observe both these 
possibilities; figure~\ref{figmpi2} is an example.
%
%
The resulting estimation of the position of the chiral~phase~transition
agrees well with those from~$\cbcex$ and~$a m_{\rm F}$.


We have also measured the other mesonic states listed in table~\ref{tabmesons}
as well as the bosonic masses~$a m_{\rm V}$~(\eq{OV}) and~$a m_{\rm S}$~(\eq{OS}).
However, these states would require much higher statistics than
we could achieve in this first explorative stage of the investigations
we are reporting here.
The mass of the composite Higgs boson $\sg$, $am_\sg$, is quite difficult to determine, which indicates that it might be rather large.
Furthermore, we can make the statement that within the (large) error bars 
we do not see from this side any indications for problems e.g.\ with
flavour symmetry restauration, i.e.\ within the error bars~$a m_\pi^{(1)}$ agrees with~$a m_\pi^{(2)}$
and~$a m_\rho^{(3)}$ with~$a m_\rho^{(4)}$, respectively.

\section{Summary and discussion}

The \SG, which we are considering as a possible substitute for the Higgs mechanism, is based on the observation that strongly coupled lattice gauge theories as a rule break dynamically a global chiral symmetry.
The model we have discussed demonstrates that, at least on the lattice, an arrangement is possible in which the Goldstone bosons and the mass of some heavy fermion arise, in a qualitative analogy to the Higgs-Yukawa sector of the standard model with a heavy quark.
The role of the scalar field is crucial, though completely different from that in the standard Higgs mechanism.

Whether a useful approach to the continuum limit can be achieved is still an open question.
However, at least in the strong coupling limit the model we have considered is promising:
In 4D it reduces to the lattice NJL~model which, though not renormalizable, has many  attributes required for a viable theory of symmetry breaking in the standard model.
In 3D and even more in 2D the lattice four fermion theories found in this limit are established as renormalizable field theories and their continuum limit is no principal obstacle.

The question whether, and exactly how, the SGM operates beyond the strong coupling limit is a difficult nonperturbative problem requiring a substantial effort. 
The difficulty is mainly due to the lack of analytic understanding of the transition between the Higgs and confinement regions.

We see essentially three possibilities.
One of them is extremely optimistic, namely that the \chup\ models could have a nontrivial fixed point, at which the cutoff can be removed completely without losing the interaction and thus the \SG.
The most obvious candidate is the point E. 
This point has been up to now investigated only in the quenched approximation. 
In the SU(2) model \cc{Bo90} some indication that the point might be nontrivial has been found, whereas in the U(1) model \cc{AlAz92,AlAz93} the data are consistent with the mean field like behaviour. 
In any case an investigation with unquenched fermions is required, which, as the experience with the strongly coupled QED shows, might be a very difficult task.

However, also if the \chup\ models would turn out to be trivial, their quantitative properties might still differ significantly from the normal Higgs-Yukawa models.
For example the upper bounds could be larger than the unitarity bounds in these models.
In contrast to the Higgs-Yukawa models there might exist a strongly interacting sector.

A somewhat disappointing possibility would be that the SGM might turn out to be equivalent to the Yukawa theory, in a similar way as the generalized four fermi theory is \cc{HaHa91,Zi91}.
Then the SGM would be only an alternative formulation of the Higgs mechanism, a particularly difficult one.
This outcome would mean that it is very difficult to substantiate the top condensate idea by means of strongly coupled gauge theories.

In any case it seems to us that an investigation of the shielded gauge mechanism of dynamical mass generation might shed new light on the little understood properties of (chiral?) gauge theories at strong coupling.

How far the SGM might be viable phenomenologically is difficult to judge at this stage.
An important question of the models replacing the Higgs mechanism is the occurence of some scalar, which might look like the Higgs boson.
Within the SGM there are two obvious candidates: the scalar bound states of 
the type $\ph^{\dagger}\ph$ and~$\cbc$. 
In the quenched approximation the first of these states scales only at the point E.
As the fermion mass scales in the vicinity of the whole NE line, one might think that the ratio of their masses might be chosen quite arbitrarily by approaching the point E in different ways.
But this must be studied in a simulation with dynamical fermions, as well as the mass of the scalar $\cbc$ state.
So at the moment we can say little about a possible occurence and mass of the Higgs-like state in the SGM.

Of course, from the phenomenological point of view, the considered lattice model is not specific enough.
It is much simpler than the existing more elaborate attempts in the continuum field theory \cc{Ho87,Hi91,Boe91,LiRo92,Ma92}.
We agree with those models only in some basic ideas.
It could be that the existing models are not yet good enough, e.g. giving incorrect top quark mass.
The contribution of the lattice methods to the further search for better models might consist in a more reliable control of the dynamical problems, treated otherwise in a rather qualitative way.

\vspace{2cm}
{\large\bf Acknowledgements}\\
We thank W. Bock, W. Franzki, P. Hasenfratz, C.B. Lang, M. Lindner, X.-Q. Luo, R.E. Shrock and M.A. Stephanov for discussions and various suggestions, and H.A. Kastrup for continuous support.
\clearpage

\bibliographystyle{wunsnot}   

\bibliography{jourabbr,our-papers,gauge,yukawa,referen}

\end{document}